\documentclass[12pt]{iopart}

\usepackage{iopams}
\usepackage[english]{babel}
\usepackage{graphicx}
\usepackage{cite}
\usepackage{color}

\begin{document}

\title[]{Quantum simulation of driven para-Bose oscillators}

\author{C. Huerta Alderete}
\address{Instituto Nacional de Astrof\'{\i}sica, \'Optica y Electr\'onica, Calle Luis Enrique Erro No. 1, Sta. Ma. Tonantzintla, Pue. CP 72840, M\'exico}

\author{B. M. Rodr\'iguez-Lara}
\address{Photonics and Mathematical Optics Group, Tecnol\'ogico de Monterrey, Monterrey 64849, Mexico. \\
Instituto Nacional de Astrof\'{\i}sica, \'Optica y Electr\'onica, Calle Luis Enrique Erro No. 1, Sta. Ma. Tonantzintla, Pue. CP 72840, M\'exico.
}
\ead{bmlara@itesm.mx}

	\begin{abstract}
	Quantum mechanics allows paraparticles with mixed Bose-Fermi statistics that have not been experimentally confirmed.
	We propose a trapped-ion scheme whose effective dynamics are equivalent to a driven para-Bose oscillator of even order.
	Our mapping suggest highly entangled vibrational and internal ion states as the laboratory equivalent of quantum simulated parabosons.
	Furthermore, we show the generation and reconstruction of coherent oscillations and para-Bose analogs of Gilmore-Perelomov coherent states from population inversion measurements in the laboratory frame.
	Our proposal, apart from demonstrating an analog quantum simulator of para-Bose oscillators, provides a quantum state engineering tool that foreshadows the potential use of paraparticle dynamics in the design of quantum information systems.
\end{abstract}


\maketitle

\section{Introduction}		
	
	The harmonic oscillator is a fundamental building block of classical and quantum physics. In quantum mechanics, Wigner found that the equations of motion do not uniquely determine the Heisenberg-Born-Jordan relation for the quantum harmonic oscillator \cite{Wigner1950p711}.
	As a result, the community started exploring deformations of the harmonic oscillator.
	One such deformation was due to the reflection operator \cite{Yang1951p788} and is known as the Calogero-Vasiliev oscillator \cite{Calogero1969p2191,Vasiliev1991p1115}. In second quantization, the commutation relations for this model,
	\begin{eqnarray}
		&\left[ \hat{A}, \hat{A}^{\dagger} \right] = 1 + (p-1) \hat{\Pi}, \nonumber \\
		&\left\{ \hat{A}, \hat{A}^{\dagger} \right\} = 2 \hat{n} +  p, \nonumber  \\
		&\left[\hat{n} , \hat{A}^{\dagger} \right] = \hat{A}^{\dagger}, \quad  \left[\hat{n} , \hat{A} \right] = - \hat{A},
	\end{eqnarray}
	deliver a paraboson algebra of order $p$ \cite{Green1953p270}.
	Here, we have used the parity operator defined as $\hat{\Pi} = e^{i \pi \hat{n}}$, the operators $\hat{n}$, $\hat{A}^{\dagger}$, and $\hat{A}$ are the number, creation and annihilation operators of the deformed oscillator, and the order parameter $p$ is a positive integer, $p \in \mathbb{Z}^{+}$.
	Note that the standard boson algebra is recovered with order $p=1$.
	Interestingly enough, quantum mechanics allows for the existence of paraparticles that have not been experimentally discovered as fundamental particles in nature \cite{Greenberg1965pB1155,Baker2015p929}.
	
	On the other hand, trapped ions have proved a reliable platform for quantum simulation, offering high precision in both parameter control and measurement \cite{Johanning2009p154009,Blatt2012p277}.
	For example, quantum simulations of relativistic \cite{Roos2011p012020} and condensed-matter \cite{Arrazola2016} physics have been realized experimentally.
	In the following, we will provide an experimental proposal involving a single trapped ion driven by two pairs of orthogonal fields tuned to the first red- and blue-sideband transitions.
	Then, we will show that a particular parameter set-up allows for the simulation of an effective model equivalent to one of the orthogonal vibrational modes coupled to the two-level ion following Jaynes-Cummings and, the other, anti-Jaynes-Cummings dynamics.
	At this point, we will demonstrate that this effective model has at least one constant of motion that allows us to map it into a driven para-Bose oscillator of even order.
	We will discuss the different regimes and related measurements that are experimentally accessible and, as an explicit example, we will focus on the generation of coherent oscillations in the paraboson number and a para-Bose analog of standard boson Gilmore-Perelomov coherent states. We will also discuss how they can be reconstructed from population inversion measurements.

\section{Experimental proposal.}

Let us start with the experimental proposal.
We consider a single trapped ion pumped by two pairs of orthogonal lasers in a configuration similar to the one we used to introduce the cross-cavity quantum Rabi model \cite{HuertaAlderete2016p},
\begin{eqnarray}
\hat{H}_{ion} &=& \frac{1}{2} \omega_{3} \hat{\sigma}_{3} + \sum_{j=1}^{2} \Bigg\{  \nu_{j} \hat{a}^{\dagger}_{j} \hat{a}_{j} + \sum_{k=-1,1} \Omega_{j,k} \times \nonumber \\
&& \times \cos \Big[ \eta_{j,k} \left( \hat{a}_{j}^{\dagger} + \hat{a}_{j} \right) - \omega_{j,k} t + \phi_{j,k} \Big] \hat{\sigma}_{j} \Bigg\}.
\end{eqnarray}
The ion is described by the transition energy $\omega_{3}$ and the Pauli matrices, $\hat{\sigma}_{j}$ with $j=1,2,3$ fulfilling the $SU(2)$ commutation relation, $\left[\hat{\sigma}_{i}, \hat{\sigma}_{j} \right] = 2i \epsilon_{ijk} \hat{\sigma}_{k}$.
The two orthogonal vibrational modes of the ion center-of-mass motion are described by the mechanical oscillation frequencies $\nu_{j}$ and the creation (annihilation) operators, $\hat{a}_{j}^{\dagger}$ ($\hat{a}_{j}$) with $j=1,2$ fulfilling the standard boson commutation relation, $\left[ \hat{a}_{j}, \hat{a}_{k}^{\dagger} \right] = \delta_{j,k}$.
Working in the Lamb-Dicke regime, $\eta_{j,k} \sqrt{\langle \hat{a}^{\dagger}_{j} \hat{a}_{j}} \rangle \ll 1$, with one of each pair of driving fields tuned to the first blue- and the first red-sideband transitions plus a small detuning,  $\omega_{j,k} = \omega_{3} + k \nu_{j} + \delta_{j,k}$ with $k= \pm 1$, we set the pump fields small detunings to be equal, $\delta_{1,k} = \delta_{2,k}=\delta_{k}$, the phases to the values $ \phi_{1,-1} = \phi_{1,1} = - \pi/2$, $\phi_{2,-1}=\pi$, and $\phi_{2,1}=0$, and tune the driving field strengths to deliver just one effective coupling strength, $ \Omega_{j,k} \eta_{j,k}e^{-\frac{1}{2}\vert \eta_{j,k}\vert^2} = g$. Then, we recover what we will call our laboratory frame Hamiltonian,
\begin{eqnarray} 
	\hat{H}_{\mathrm{Lab}} &=& \frac{1}{2} \omega_{0} \hat{\sigma}_{3} + \omega \left( \hat{a}^{\dagger}_{1} \hat{a}_{1} + \hat{a}^{\dagger}_{2} \hat{a}_{2}\right)  + g \left[ \left(\hat{a}^{\dagger}_{1} + \hat{a}_{1} + \hat{a}^{\dagger}_{2} - \hat{a}_{2}\right)\hat{\sigma}_{+} + \mathrm{h.c.} \right],
\end{eqnarray}
where the effective qubit frequency is given by the halved addition of small detunings $\omega_{0} = - \left( \delta_{-1} + \delta_{1} \right)/2$ and the effective field frequencies by their halved difference, $\omega = \left( \delta_{-1} - \delta_{1} \right)/2$.
This is a cross-cavity quantum Rabi model type Hamiltonian where the fields have equal frequencies and couple with the same strength to the qubit \cite{Chilingaryan2015p245501,HuertaAlderete2016p}.
It is well-known that it can be written as a Hamiltonian model where the ion is coupled under Jaynes-Cummings and anti-Jaynes-Cummings dynamics to each of the fields with identical effective coupling strength \cite{HuertaAlderete2016p},
\begin{eqnarray} 
	\hat{H}_{cc} &=& \frac{1}{2}\omega_{0} \hat{\sigma}_{3} +   \sum_{j=1}^{2} \omega \hat{a}^{\dagger}_{j} \hat{a}_{j} + \sqrt{2} g \left[ \left( \hat{a}^{\dagger}_{1} - \hat{a}_{2} \right) \hat{\sigma}_{+} + \left( \hat{a}_{1} - \hat{a}^{\dagger}_{2} \right) \hat{\sigma}_{-}\right], \label{eq:HJCaJC}
\end{eqnarray}
after using Schwinger two-boson representation of $SU(2)$ to effect a rotation $\hat{D}_{y}(\theta) = e^{i \theta \hat{J}_{2} }$ with $\theta = \pi/2$ and $ \hat{J}_{2} = -\frac{i}{2} \left( \hat{a}_{1}^{\dagger}\hat{a}_{2}-\hat{a}_{1}\hat{a}_{2}^{\dagger}\right)$,  
\begin{eqnarray}
	\hat{D}_{y}(\theta) \hat{a}_{1} \hat{D}_{y}^{\dagger}(\theta) &=& \cos \left(\frac{\theta}{2}\right) \hat{a}_{1} - \sin\left(\frac{\theta}{2}\right)\hat{a}_{2}, \nonumber \\ 
	\hat{D}_{y}(\theta) \hat{a}_{2} \hat{D}_{y}^{\dagger}(\theta) &=& \cos \left(\frac{\theta}{2}\right) \hat{a}_{2} + \sin\left(\frac{\theta}{2}\right)\hat{a}_{1}.
\end{eqnarray}
This new effective Hamiltonian conserves the scaled difference between population inversions from the two $SU(2)$ representations,
\begin{eqnarray}
	\hat{\eta}_{cc} =  -2\hat{J}_{3} + \frac{1}{2} \left( \hat{\sigma}_{3} + 1 \right),
\end{eqnarray}
where the effective population inversion in the two-boson representation of $SU(2)$ is given by
$\hat{J}_{3} = \frac{1}{2} \left(\hat{a}^{\dagger}_{1}\hat{a}_{1} - \hat{a}^{\dagger}_{2}\hat{a}_{2} \right) $.
Note that this operator is composed by the excitation number from the Jaynes-Cummings and anti-Jaynes-Cummings models \cite{RodriguezLara2005p023811}. In the laboratory frame,
this operator,
\begin{eqnarray}
	\hat{\eta}_{\mathrm{Lab}} = - 2 \hat{J}_{1} + \frac{1}{2} \left( \hat{\sigma}_{3} + 1 \right),
\end{eqnarray}
is related to the population inversion of just the qubit and the mixing rate of the vibrational modes,	$\hat{J}_{1} = \frac{1}{2} \left( \hat{a}_{1}^{\dagger}\hat{a}_{2} + \hat{a}_{1}\hat{a}_{2}^{\dagger}\right)$, and, of course, it is a conserved variable, $\left[ \hat{\eta}_{x} , \hat{H}_{x}\right]=0$ with $x= \mathrm{Lab}, cc$.

As a final note on the experimental realization, we want to emphasize the practical potential of our experimental proposal within the limitations provided by state-of-the-art techniques available for precision laser frequency and intensity tuning for the four driving lasers and controlling the decoherence of the two-vibrational modes.
	
\section{Diagonalization in the qubit basis}
	In order to isolate the field dynamics, we can move into a frame defined by the population inversion difference $\hat{\eta}_{cc}$ rotating at the qubit frequency $\omega_{0}$, implement a rotation of $\pi/4$ around $\hat{\sigma}_{2}$, and, then,  diagonalize the resulting Hamiltonian in the qubit basis using the Fulton-Gouterman approach \cite{Moroz2016p50004},
	\begin{eqnarray}
	\hat{H}_{FG} &=& \hat{U}_{FG} \hat{H} \hat{U}_{FG}^{\dagger} \nonumber \\
	&=&  \hat{H}_{+} \vert e \rangle \langle e \vert + \hat{H}_{-} \vert g \rangle \langle g \vert,
	\end{eqnarray}
where the Fulton-Gouterman transform is the following,
	\begin{eqnarray}
	\hat{U}_{FG} = \frac{1}{\sqrt{2}} \left( \begin{array}{cc} 1 & \hat{\Pi}_{12} \\ 1 & - \hat{\Pi}_{12} \end{array} \right),
	\end{eqnarray}
with the two fields parity defined as $\hat{\Pi}_{12} = e^{i \pi \left( \hat{a}_{1}^{\dagger} \hat{a}_{1} + \hat{a}_{2}^{\dagger} \hat{a}_{2}   \right) } $, such that the effective field Hamiltonians,
	\begin{eqnarray}
	\hat{H}_{\pm} &=& \sum_{j=1}^{2} \left[ \omega + (-1)^{j} \omega_{0} \right] \hat{a}_{j}^{\dagger} \hat{a}_{j}  - \frac{(-1)^{j}}{\sqrt{2}} g \times \nonumber \\
&&	 \times \left\{  \hat{a}_{j} \left[ 1 \pm (-1)^{j} \hat{\Pi}_{12} \right] + \left[ 1 \pm (-1)^{j}\hat{\Pi}_{12} \right] \hat{a}_{j}^{\dagger} \right\},
	\end{eqnarray}
	describe two boson fields interacting through a nonlinear coupling that depends on the total parity.
	In summary, the mapping from the laboratory to the Fulton-Gouterman frame is provided by the unitary transformation,
	\begin{eqnarray}
	\hat{T}_{Lab \rightarrow FG} =   \hat{U}_{FG} e^{i \frac{\pi}{4} \hat{\sigma}_{2}} e^{i \omega_{0} \hat{\eta}_{cc} t}   e^{i \theta \hat{J}_{2} }.
	\end{eqnarray}
	The conserved operator in the Fulton-Gouterman frame becomes a different scaled difference of population inversions of the two $SU(2)$ representations,
	\begin{eqnarray}
	\hat{\eta}_{FG} = \hat{\eta}_{+} \vert e \rangle \langle e \vert + \hat{\eta}_{-} \vert g \rangle \langle g \vert.
	\end{eqnarray}
	where the operators associated to the excited and ground state diagonal components are diagonal themselves,
	\begin{eqnarray}
    \hat{\eta}_{\pm} = -2\hat{J}_{3} + \frac{1}{2}\left(1\mp\hat{\Pi}_{12} \right)
	\end{eqnarray}
	such that, obviously, these pure field operators commute with their respective accompanying field Hamiltonian, $\left[\hat{H}_{\pm}, \hat{\eta}_{\pm}\right]=0$.
	Thus, we can set ourselves to partition the whole Hilbert space into subspaces that keep the average of these operators constant.
	
In order to isolate the field dynamics, we can move into a frame defined by the population inversion difference $\hat{\eta}_{cc}$ rotating at the qubit frequency $\omega_{0}$, implement a rotation of $\pi/4$ around $\hat{\sigma}_{2}$, and, then,  diagonalize the resulting Hamiltonian in the qubit basis using the Fulton-Gouterman approach \cite{Moroz2016p50004},
\begin{eqnarray}
	\hat{H}_{FG} &=& \hat{U}_{FG} \hat{H} \hat{U}_{FG}^{\dagger} \nonumber \\
	&=&  \hat{H}_{+} \vert e \rangle \langle e \vert + \hat{H}_{-} \vert g \rangle \langle g \vert, \label{eq:FGHam}
\end{eqnarray}
where the Fulton-Gouterman transform is the following,
\begin{eqnarray}
	\hat{U}_{FG} = \frac{1}{\sqrt{2}} \left( \begin{array}{cc} 1 & \hat{\Pi}_{12} \\ 1 & - \hat{\Pi}_{12} \end{array} \right),
\end{eqnarray}
with the two fields parity defined as $\hat{\Pi}_{12} = e^{i \pi \left( \hat{a}_{1}^{\dagger} \hat{a}_{1} + \hat{a}_{2}^{\dagger} \hat{a}_{2}   \right) } $, such that the effective field Hamiltonians,
\begin{eqnarray}
	\hat{H}_{\pm} &=& \sum_{j=1}^{2} \left[ \omega + (-1)^{j} \omega_{0} \right] \hat{a}_{j}^{\dagger} \hat{a}_{j}  - \frac{(-1)^{j}}{\sqrt{2}} g \times \nonumber \\
	&&	 \times \left\{  \hat{a}_{j} \left[ 1 \pm (-1)^{j} \hat{\Pi}_{12} \right] + \left[ 1 \pm (-1)^{j}\hat{\Pi}_{12} \right] \hat{a}_{j}^{\dagger} \right\}, \label{eq:EFHam}
\end{eqnarray}
describe two boson fields interacting through a nonlinear coupling that depends on the total parity.
In summary, the mapping from the laboratory to the Fulton-Gouterman frame is provided by the unitary transformation,
\begin{eqnarray}
	\hat{T} =   \hat{U}_{FG} e^{i \frac{\pi}{4} \hat{\sigma}_{2}} e^{i \omega_{0} \hat{\eta}_{cc} t}   e^{i \frac{\pi}{2} \hat{J}_{2} }.
\end{eqnarray}
Thus, states and operators can be transformed from the laboratory frame to the Fulton-Gouterman frame, $\vert \psi \rangle_{FG} = \hat{T} \vert \psi \rangle_{Lab}$ and $\hat{O}_{FG} = \hat{T} \hat{O}_{Lab} \hat{T}^{\dagger}$, and vice-versa, $\vert \psi \rangle_{Lab} = \hat{T}^{\dagger} \vert \psi \rangle_{FG}$ and $\hat{O}_{Lab} = \hat{T}^{\dagger} \hat{O}_{FG} \hat{T}$.
Following this prescription, the conserved operator in the Fulton-Gouterman frame becomes a different scaled difference of population inversions of the two $SU(2)$ representations,
\begin{eqnarray}
	\hat{\eta}_{FG} = \hat{\eta}_{+} \vert e \rangle \langle e \vert + \hat{\eta}_{-} \vert g \rangle \langle g \vert.
\end{eqnarray}
where the operators associated to the excited and ground state diagonal components are diagonal themselves,
\begin{eqnarray}
	\hat{\eta}_{\pm} = -2\hat{J}_{3} + \frac{1}{2}\left(1\mp\hat{\Pi}_{12} \right)
\end{eqnarray}
such that, obviously, these pure field operators commute with their respective accompanying field Hamiltonian, $\left[\hat{H}_{\pm}, \hat{\eta}_{\pm}\right]=0$.
Thus, we can set ourselves to partition the whole Hilbert space into subspaces that keep the average of these operators constant.

\section{Partition of the Hilbert space}

The system Hamiltonian diagonalized in the qubit basis, $\hat{H}_{FG}$, allows us to focus on the diagonalization of just the effective field Hamiltonians, $\hat{H}_{\pm}$.
Starting from the vacuum state and verifying the action of these effective field Hamiltonians on it, we can partition the corresponding Hilbert spaces,
\begin{eqnarray}
	\mathcal{H}_{\pm} = \bigoplus_{j=0}^{\infty}  \mathcal{H}_{\pm, j},
\end{eqnarray}
where the subspaces corresponding to the effective field Hamiltonian accompanying the upper diagonal element are spanned by the following orthonormal bases, 
\begin{eqnarray}
	\mathcal{H}_{+,2N} &=& \left\{ \vert +, 2N;  k \rangle ~\mid~  \vert +, 2N;  k \rangle \equiv \vert h(k+4N+1), h(k) \rangle \right\}, \nonumber \\
	\mathcal{H}_{+,2N+1} &=& \left\{ \vert +, 2N+1;  k \rangle ~\mid~  \vert +, 2N+1;  k \rangle \equiv \vert h(k), h(k+4N+3) \rangle \right\}, \nonumber \\
\end{eqnarray}
and the subspaces corresponding to the lower diagonal effective field Hamiltonian, 
\begin{eqnarray}
	\mathcal{H}_{-,2N} &=& \left\{ \vert -, 2N;  k \rangle ~\mid~  \vert -, 2N;  k \rangle \equiv \vert h(k), h(k+4N+3) \rangle \right\}, \nonumber \\
	\mathcal{H}_{-,2N+1} &=& \left\{ \vert -, 2N+1;  k \rangle ~\mid~  \vert -, 2N+1;  k \rangle \equiv  \vert h(k+4N+1), h(k) \rangle \right\}, \nonumber \\
\end{eqnarray}
with $N,k=0,1,2\ldots$, we used the shorthand notation $\vert m, n \rangle \equiv \vert m \rangle_{1} \vert n \rangle_{2}$ for the states related to the two vibrational modes, and the following integer valued function,
\begin{eqnarray}
	h(k) = \frac{1}{4} \left( 2k - 1 + e^{i\pi k} \right)
\end{eqnarray}
Note that the subspaces $\mathcal{H}_{+,2N}$ and $\mathcal{H}_{-,2N+1}$, as well as $\mathcal{H}_{+,2N+1}$ and $\mathcal{H}_{-,2N}$, are spanned by identical orthonormal bases but they are associated to different qubit sectors in the Fulton-Gouterman frame, $\hat{H}_{FG}$. 
We will later identify the shorthand notation $\vert \pm, N; k \rangle$ as the $k$-th Fock state of the para-Bose oscillator of order $p=2(N+1)$ related to the positive and negative subspaces defined by the diagonalization in the qubit basis. 

We can readily identify orthonormal basis components, $ \vert +, j; k \rangle \vert e \rangle$ and $\vert -, j; k \rangle \vert g \rangle$, in the Fulton-Gouterman frame as eigenstates of Schwinger two-boson operator $\hat{J}_{1}$ in the laboratory frame, 
\begin{eqnarray}
	\hat{J}_{1} \hat{T}^{\dagger} \vert +,j;k \rangle \vert e \rangle &=& \frac{e^{i \pi j}}{4}  \left[ 2 j + \left(1 - e^{i \pi k}\right) \right] \hat{T}^{\dagger} \vert +, j;k \rangle \vert e \rangle , \nonumber \\
	\hat{J}_{1} \hat{T}^{\dagger} \vert - ,j;k \rangle \vert g \rangle &=& \frac{e^{i \pi (j + 1)}}{4}  \left[ 2 \left(j + e^{i \pi j}\right) + \left(1 - e^{i \pi k}\right) \right] \hat{T}^{\dagger} \vert -,j;k \rangle \vert g \rangle ,
\end{eqnarray}
for each and every representation defined by the total number of bosons in the vibrational modes.
We can also verify the expectation value of our difference in the population inversion difference, $\hat{\eta}_{\pm}$, for these orthonormal bases,
\begin{eqnarray} \nonumber
	\langle +, 2N; k \vert ~\hat{\eta}_{+} ~\vert +, 2N; k \rangle &=& -2N, \nonumber \\
	\langle +, 2N+1; k \vert ~\hat{\eta}_{+} ~\vert +, 2N+1; k \rangle &=& 2\left( N+1 \right), \nonumber \\
	\langle -, 2N; k \vert ~\hat{\eta}_{+} ~\vert -, 2N; k \rangle &=& 2\left( N+1 \right), \nonumber \\
	\langle -, 2N; k \vert ~\hat{\eta}_{+} ~\vert -, 2N+1; k \rangle &=& -2N,
\end{eqnarray}
and find that it is conserved,
\begin{eqnarray}
	\mathcal{H}_{+,2N},\mathcal{H}_{-,2N+1} &:& \quad \langle \hat{\eta}_{\pm} \rangle = - 2 N, \nonumber  \\
	\mathcal{H}_{+,2N+1},\mathcal{H}_{-,2N}&:& \quad \langle \hat{\eta}_{\pm} \rangle = 2 (N+1).
\end{eqnarray}
Note that the difference of population inversions of the two $SU(2)$ representations in the Fulton-Gouterman frame, $\hat{\eta}_{FG}$, is twofold degenerate for these bases,
\begin{eqnarray}
\hspace{-1.5cm}	\langle e \vert \langle +, 2N; k \vert ~\hat{\eta}_{FG} ~\vert +, 2N; k \rangle \vert e \rangle &=& \langle g \vert \langle -, 2N+1; k \vert ~\hat{\eta}_{FG} ~\vert -, 2N+1; k \rangle \vert g \rangle, \nonumber \\
\hspace{-1.5cm}\langle e \vert \langle +, 2N+1; k \vert ~\hat{\eta}_{FG} ~\vert +, 2N+1; k \rangle \vert e \rangle &=& \langle g \vert \langle -, 2N; k \vert ~\hat{\eta}_{FG} ~\vert -, 2N; k \rangle \vert g \rangle.
\end{eqnarray}

\section{Generalized Para-Bose oscillator}	
We can use these orthonormal bases to project the auxiliary field Hamiltonians, $\hat{H}_{\pm}$, into these subspaces and realize they can be written as driven nonlinear oscillators,
\begin{eqnarray}
	\hat{H}_{\pm, N} &=& \omega\left(\hat{n}_{N} + N\right) \mp \frac{\omega_{0}}{2} e^{i\pi \left(\hat{n}_{N}+N\right)} \pm g \left( \hat{A}_{N}^{\dagger} + \hat{A}_{N} \right) + \lambda_{\pm, N},
\end{eqnarray}
where the auxiliary constants are given in the following,
\begin{eqnarray}
	\lambda_{+,N} & = & \omega_{0}\left( N + \frac{1}{2}\right)e^{i\pi N}, \nonumber \\
	\lambda_{-,N} & = & \left[\omega - \omega_{0}\left(N+\frac{1}{2}\right)\right]e^{i\pi N}-\omega_{0}.
\end{eqnarray}
and we have defined the creation and annihilation operators,
\begin{eqnarray}
	\hat{A}_{N}^{\dagger} &=& \hat{a}^{\dagger}_{N}~ f_{N}(\hat{n}_{N}), \nonumber \\
	\hat{A}_{N} &=& f_{N}(\hat{n}_{N}) ~ \hat{a}_{N},
\end{eqnarray}
given in terms of a nonlinear deformation function,
\begin{eqnarray}
	f_{N}(k) = \sqrt{\frac{2k + \left( 2N + 3 \right) + \left( 2N + 1 \right)e^{i\pi k}}{2(k + 1)}} ~e^{i \pi (k + N)},
\end{eqnarray}
and the standard boson operators for each subspace.
We can calculate the actions of the nonlinear operators on the Fock states of each subspace,
\begin{eqnarray}
	\hat{A}_{N}^{\dagger} \vert \pm,N; k \rangle & = & \sqrt{k+1} ~ f_{N}(k)~ \vert \pm, N; k + 1 \rangle, \nonumber \\
	\hat{A}_{N} \vert \pm,N; k \rangle & = & \sqrt{k} ~ f_{N}(k-1) ~ \vert \pm, N; k - 1 \rangle, \nonumber \\
	\hat{n}_{N} \vert \pm, N; k \rangle & = & k ~\vert \pm,N; k \rangle,
\end{eqnarray}
and, most important at this point, the action of this particular combination on the vacuum state of each subspace,
\begin{eqnarray}
	\hat{A}_{N} \hat{A}_{N}^{\dagger} \vert \pm, N;0\rangle = 2(N+1) \vert \pm,N;0\rangle,
\end{eqnarray}
suggest that each subspace corresponds to a paraparticle Hilbert space  of even order $p=2(N+1)$ \cite{Green1953p270}.
Note that our quantum simulation will never provide ordinary bosons as they have odd order $p=1$.
It is straightforward to check that we have an even order para-Bose algebra in our hands,
\begin{eqnarray}
	&\left[ \hat{A}_{N}, \hat{A}^{\dagger}_{N} \right] = 1 + (2N + 1) \hat{\Pi}_{N}, \nonumber \\
	&\left\{ \hat{A}_{N}, \hat{A}^{\dagger}_{N} \right\} = 2 \hat{n}_{N} + 2(N +1), \nonumber  \\
	&\left[\hat{n}_{N} , \hat{A}^{\dagger}_{N} \right] = \hat{A}^{\dagger}_{N}, \quad  \left[\hat{n}_{N} , \hat{A}_{N} \right] = - \hat{A}_{N},
\end{eqnarray}
This helps us realize that the auxiliary field Hamiltonians,
\begin{eqnarray}
	\hat{H}_{\pm, N} = \frac{\omega}{2} \left\{ \hat{A}_{N} \pm \frac{g}{\omega}, \hat{A}_{N}^{\dagger} \pm \frac{g}{\omega}  \right\} + F_{\pm, N} (\hat{n}_{N}), \label{eq:AuxFHam}
\end{eqnarray}
are nothing else than displaced para-Bose oscillators plus an additional diagonal term that depends only on the parity,
\begin{eqnarray}
	F_{\pm, N}(\hat{n}_{N}) = \lambda_{\pm, N} - \omega\left(1+\frac{g^{2}}{\omega^{2}}\right) \mp \frac{1}{2} \omega_{0} e^{i \pi (\hat{n}_{N} + N)} .
\end{eqnarray}
We hope our proposal encourages further quantum simulations of parabose oscillators of odd order, para-Fermi oscillators, or even deformations of the Jaynes-Cummings model where standard bosons are replaced by Calogero-Vasiliev oscillators \cite{Crnugelj1994p3545}.

\section{Driven parabose oscillator dynamics}
\begin{figure}[htbp]
	\centering
	\includegraphics{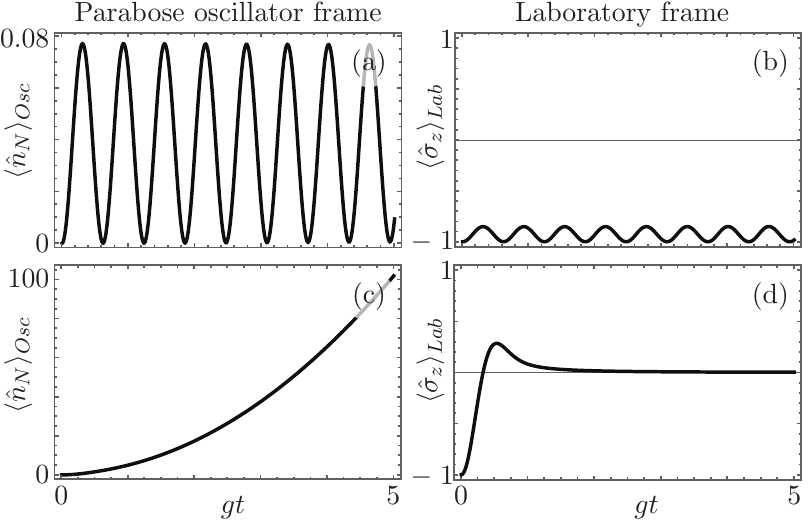}
	\caption{\label{fig:Fig1}Time evolution of the mean para-Bose number operator in the quantum simulation frame, $\langle \hat{n}_{N} \rangle_{Osc}$, and corresponding population inversion in the laboratory frame, $\langle \hat{\sigma}_{z} \rangle_{Lab}$, for a starting $p=2(N+1)=2$ para-Bose vacuum state under weak coupling, (a),(b) $g= 0.1 \omega$, and pure pumping, (c),(d) $\omega=0$, dynamics.}
\end{figure}

Note that in the case of null effective qubit frequency, $\omega_{0} = 0$, that is small driving field detunings $\delta_{-1}= - \delta_{1}$, we recover a  driven para-Bose oscillator of even order $p=2(N+1)$,
\begin{eqnarray}
	\hat{H}_{Osc} = \omega \hat{n}_{N} + g \left( \hat{A}_{N} + \hat{A}_{N}^{\dagger} \right).
\end{eqnarray}
Note that this effective driven para-Bose oscillator can be constructed from either one of the projected auxiliary field Hamiltonians, $\hat{H}_{\pm, N}$.
We will now study this model in order to provide a particular example and create some intuition.
In the standard boson case, which is not covered by our quantum simulation, it is straightforward to identify two extremal cases, one for free evolution, $g=0$, where an initial Fock state will only gather a phase proportional to the propagation time; in our simulation this regime does not make sense because it implies the absence of driving fields.
And the other for pure driving, $\omega = 0$, where an initial Fock state will become a displaced number state due to the continuous pumping; this can be explored for parabosons with our quantum simulation.
Another interesting regime for standard bosons occurs where the coupling is small with respect to the field frequency, $g \ll \omega$; here an initial Fock state will show coherent oscillations and also can be explored with our quantum simulation.
These behaviors for standard bosons are easily visualized in the so-called Glauber-Fock oscillator from photonic waveguide arrays  \cite{RodriguezLara2011p053845,VillanuevaVergara2015p}.
In Fig. \ref{fig:Fig1} we show that a similar type of dynamics occur in the driven para-Bose oscillator for the lowest order $p=2$, that is $N=0$. For example, let us consider as initial state of the oscillator the vacuum state of the subspace $\mathcal{H}_{+,0}$, that is $\vert +,0;0\rangle = \vert 0,0 \rangle$, which in the Fulton-Gouterman frame corresponds to the state $ \vert \psi(0) \rangle_{FG} = \vert 0,0,e \rangle_{FG}$, then, in the laboratory frame this parabose vacuum state of order $p=2$ means preparing the following state: 
\begin{eqnarray}
	\vert \psi(0) \rangle_{Lab} &=& \hat{T}^{\dagger} \vert 0,0,e \rangle_{FG}, \nonumber \\
	&=& \vert 0, 0, g \rangle_{Lab}.
\end{eqnarray}
In other words, in the laboratory frame, we need to cool down the vibrational modes to the lowest mechanical state. This is experimentally feasible following standard ion-trap state engineering protocols for each vibrational mode separately \cite{Law1996p1055}.
Note that this state is an eigenstate of pure Jaynes-Cummings dynamics.
This initial state shows coherent oscillations of the mean paraboson number in the small coupling regime with $g = 0.1 \omega$, Fig. \ref{fig:Fig1}(a).
These coherent oscillations double the amplitude and have a slightly larger frequency than  those obtained in the standard boson Glauber-Fock oscillator for exactly the same parameters.
The mean paraboson number oscillation in the quantum simulation frame translate to oscillation of the population inversion in the laboratory frame, Fig. \ref{fig:Fig1}(b), because the latter is related to the parity of the para-Bose state, which is given explicitly in the following for the case at hand: 
\begin{eqnarray}
	\langle \hat{\sigma}_{z}(t) \rangle_{Lab} &=&  \langle \psi(0) \vert  \hat{U}^{\dagger}(t) \hat{\sigma}_{z} \hat{U}(t) \vert \psi(0) \rangle_{Lab}, \nonumber \\
	&=& - \langle 0,0\vert e^{i H_{+} t} \hat{\Pi}_{12}  e^{-i H_{+} t} \vert 0,0 \rangle_{FG}  \langle e \vert \hat{\sigma}_{z} \vert e \rangle_{FG}, \nonumber  \\
	&=& - \langle \psi(t) \vert e^{i \pi \hat{n}_{N}} \vert \psi(t) \rangle_{Osc}.
\end{eqnarray}
Here, we have used the fact that the population inversion in the laboratory frame transforms into a paritylike operator in the Fulton-Gouterman frame, $\hat{T} \hat{\sigma}_{z} \hat{T}^{\dagger} = -\hat{\sigma}_{z} e^{i \pi \left( \hat{a}_{1}^{\dagger} \hat{a}_{1} + \hat{a}_{2}^{\dagger} \hat{a}_{2}\right)}$, and the time evolution acting onto the initial state becomes the evolution provided by just one of the auxiliary field Hamiltonians,  $\hat{T} \hat{U}^{\dagger}(t) \hat{T}^{\dagger} \vert 0,0,e \rangle_{FG} = e^{-i H_{+} t} \vert 0,0,e \rangle_{FG}$.

We can generalize this result and realize that, for pure driving, $\omega = 0$, the initial para-Bose vacuum state becomes the analog of a standard boson Gilmore-Perelomov coherent state,
\begin{eqnarray}
	\vert \pm, N; \beta \rangle &=& e^{-i g (\hat{A}^{\dagger}_{N} + \hat{A}_{N}) t} \vert \pm, N; 0 \rangle, 
\end{eqnarray}
and for the para-Bose oscillator at hand, $N=0$, we can calculate the following form:
\begin{eqnarray}
	\hspace{-1.5cm} \vert \pm, 0; \beta \rangle &=& e^{-i g (\hat{A}^{\dagger}_{N} + \hat{A}_{N}) t} \vert \pm, 0; 0 \rangle, \nonumber \\
	&=& \sum_{j=0}^{\infty} \frac{j!}{(2j)!}(\sqrt{2} g t)^{2 j} ~_{1}F_{1}\left( j+1, j+\frac{1}{2}; -\frac{1}{2} g^{2} t^{2}\right) \vert \pm, 0; 2j \rangle + \nonumber \\
	&&- i \sum_{j=0}^{\infty} \frac{j! \sqrt{j+1}}{(2j+1)!} (\sqrt{2} g t)^{2 j + 1} ~_{1}F_{1}\left( j+2, j+\frac{3}{2}; -\frac{1}{2} g^{2} t^{2}\right) \vert \pm, 0; 2j+1 \rangle, \nonumber \\
\end{eqnarray}
where we have used the notation $_{1}F_{1}(a,b;z)$ for the confluent hypergeometric function, with coherent parameter equal to the scaled time, $\beta = - i g t$.
Also, in this frame, it is cumbersome but possible to provide a closed form expression for the expectation value of the number operator for these para-Bose analogs of Gilmore-Perelomov coherent states,
\begin{eqnarray}
	\langle \hat{n} \rangle_{Osc} &=& \frac{g^2 t^2}{2} \left[ 1 + {}~_{2}F_{2}\left(1,1; \frac{3}{2},2;-2 g^2 t^2\right) \right] , \quad N=0,
\end{eqnarray}
where the function ${}_{2}F_{2}(a_{1}, a_{1}; b_{1}, b_{2}; z)$ is the generalized hypergeometric function, Fig. \ref{fig:Fig1}(c).
In the laboratory frame, we can write this para-Bose analog of a coherent state as a nonseparable state, for the sake of space we will write it one step before the laboratory frame,
\begin{eqnarray}
\hspace{-1.5cm}	\vert +,0; \beta \rangle_{cc} &=& e^{-i \frac{\pi}{4} \hat{\sigma}_{2}} \hat{U}_{FG}^{\dagger} \vert \pm, 0; \beta \rangle \vert e \rangle, \nonumber \\
&=&\sum_{j=0}^{\infty} \frac{j!}{(2j)!}(\sqrt{2} g t)^{2 j} ~_{1}F_{1}\left( j+1, j+\frac{1}{2}; -\frac{1}{2} g^{2} t^{2}\right) \vert j,j,g \rangle + \nonumber \\
	&&- i \sum_{j=0}^{\infty} \frac{j! \sqrt{j+1}}{(2j+1)!} (\sqrt{2} g t)^{2 j + 1} ~_{1}F_{1}\left( j+2, j+\frac{3}{2}; -\frac{1}{2} g^{2} t^{2}\right) \vert j+1,j,e \rangle, \nonumber \\
\end{eqnarray}
It is possible to confirm numerically that these states have balanced even and odd parity components for large coherent parameters.
This could be seen in the laboratory frame through the population inversion, Fig. \ref{fig:Fig1}(d).
Note that we have focus on just the time evolution of the vacuum state of para-Bose particles of order $p=2$, but any other initial state of higher-order para-Bose oscillators might be experimentally attainable with current state engineering techniques \cite{Drobny1998p2481}.
With this in mind, the proposed model might be used as a quantum state engineering process that delivers a peculiar tripartite entangled state of two vibrational modes and a qubit  by simple time evolution, which might be of use for quantum information processing.

\begin{figure}[htbp]
	\centering
	\includegraphics{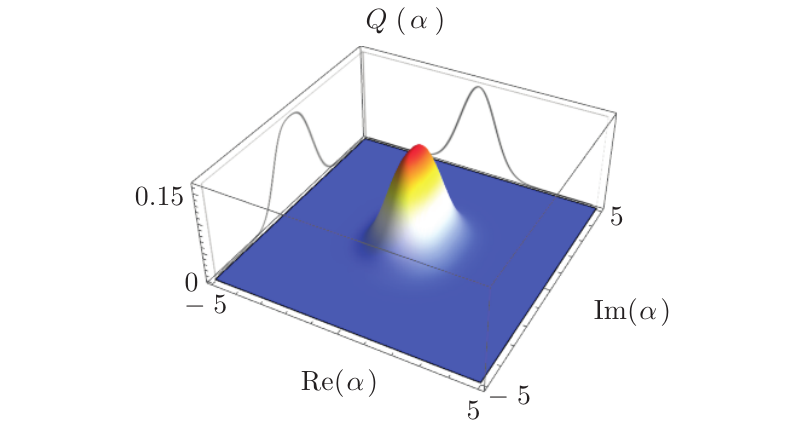}
	\caption{\label{fig:Fig2} Husimi Q function in the laboratory frame for the reduced density matrix of the first vibrational mode of the para-Bose analog of a Gilmore-Perelomov coherent state of order $p=2$ and coherent parameter $\beta= i g t$, $\vert +, 0; \beta \rangle$, obtained via evolution of the para-Bose vacuum state of order $p=2$ under purely pumped dynamics, $\omega = 0$, at the scaled time $g t = 1$.}
\end{figure}

In trapped ion experiments, it is feasible to reconstruct the Wigner function of single mode vibrational states from population inversion measurements \cite{Poyatos1996p1966,DHelon1996p25,Leibfried1996p01713}.
Thus, a viable single vibrational mode reconstruction scenario would initialize the system, let it evolve under the quantum simulation, stop the simulation, turn off the interaction between inner and vibrational modes, and turn on the protocol to recover phase space information of a single vibrational mode.
Then, the whole process can be repeated for the other vibrational mode. 
Numerically, it is simpler to calculate Hussimi Q function \cite{Gerry2005}, 
\begin{eqnarray}
	Q(\alpha) = \frac{1}{\pi} \langle  \alpha \vert \hat{\rho}_{f} \vert \alpha \rangle,
\end{eqnarray}
which can be reconstructed from the experimental Wigner function, where the parameter $\alpha$ is a complex number, and we have used the notation $\vert \alpha \rangle$ for standard boson coherent states and $\hat{\rho}_{f}$ for the reduced density matrix of the single vibrational mode.
For example, if we consider the pure pumping scheme, $\omega=0$, where para-Bose analogs of standard boson coherent states are generated, we can numerically calculate the Husimi Q function of the first vibrational mode in the laboratory frame and realize that it presents quadrature squeezing; see Fig. \ref{fig:Fig2}. 
We will cover the properties of these para-Bose analogs of standard boson Gilmore-Perelomov coherent states in detail somewhere else.

\section{Conclusion}
In summary, we have proposed a trapped ion configuration where the interaction of the ion internal degree of freedom with two orthogonal center-of-mass motion degrees of freedom can be reduced to that of an even order para-Bose oscillator.
We have discussed in detail both the experimental proposal and the analytic approach that diagonalizes the system for the internal degree of freedom and, then, partitions the Hilbert space for the vibrational degrees of freedom delivering Hilbert subspaces corresponding to even order parabosons.
As a particular example, we have focused on a driven para-Bose oscillator and demonstrated that it can produce coherent oscillations of the mean paraboson number and a para-Bose analog of Gilmore-Perelomov coherent states that can be reconstructed in the laboratory frame through population inversion measurements.
These dynamics might be of use in the quantum engineering of tripartite entangled states.

	\section*{Acknowledgments}
	C.H.A. acknowledges financial support from CONACYT doctoral grant $\#455378$ and B.M.R.L. from CONACYT CB-2015-01 project $\#255230$.
	 The authors thank Changsuk Noh for fruitful discussions.

	

\end{document}